\documentclass[prl,aps,10pt, twocolumn, superscriptaddress, floatfix, nofootinbib, amsmath, amssymb, preprintnumbers]{revtex4-1}

\usepackage{dcolumn}
\usepackage{verbatim}

\usepackage{breakurl}

\usepackage{lmodern}

\pdfoutput=1
\usepackage{graphicx}
\usepackage{bm}
\usepackage{amsmath}
\usepackage{amsfonts}
\usepackage{amssymb}
\usepackage{multirow}
\usepackage[colorlinks,linktocpage,linkcolor=cyan,citecolor=cyan]{hyperref}
\usepackage{adjustbox}
\usepackage{array}
\usepackage[capitalise]{cleveref}
\usepackage{acro}
\usepackage[utf8]{inputenc}
\usepackage{CJKutf8}
\usepackage{amssymb}
\usepackage{subfigure}

\usepackage{tikz}
\usepackage{pgfplots}
\pgfplotsset{compat=newest,every axis plot/.append style={line width=1pt}}

\crefname{figure}{Fig.}{Figs.}
\Crefname{figure}{Fig.}{Figs.}
\def\({\left(}
\def\){\right)}
\def\[{\left[}
\def\]{\right]}

\newcommand{\be}{{\begin{eqnarray}}}
\newcommand{\ee}{{\end{eqnarray}}}

\newcommand{\overbar}[1]{\mkern 1.5mu\overline{\mkern-1.5mu#1\mkern-1.5mu}\mkern 1.5mu}

\newcommand{\fnl}{f_\mathrm{nl}}

\newcommand{\cA}{\mathcal{A}}

\newcommand{\cH}{\mathcal{H}}
\newcommand{\cS}{\mathcal{S}}
\newcommand{\cO}{\mathcal{O}}

\newcommand{\bk}{\mathbf{k}}

\newcommand{\bq}{\mathbf{q}}
\newcommand{\bp}{\mathbf{p}}
\newcommand{\bx}{\mathbf{x}}

\newcommand{\ud}{\mathrm{d}}

\newcommand{\uGW}{\mathrm{gw}}

\newcommand{\Beq}{\begin{align}}
\newcommand{\Eeq}{\end{align}}

\DeclareAcronym{BH}{
  short = BH ,
  long = black hole ,
  short-plural = s ,
}
\DeclareAcronym{SNR}{
  short = SNR ,
  long = signal-to-noise ratio ,
  short-plural = s ,
}
\DeclareAcronym{IMRPPv2}{
  short = ,
  long = {\normalsize IMRP}{\footnotesize HENOM}{\normalsize P}v2 ,
  short-plural = ,
}
\DeclareAcronym{SFR}{
  short = SFR ,
  long = star formation rate ,
  short-plural =  ,
}
\DeclareAcronym{IMR}{
  short = IMR ,
  long = inspiral-merger-ringdown ,
  short-plural =  ,
}
\DeclareAcronym{ABH}{
	short = ABH ,
	long  = astrophysical black hole,
  short-plural = s ,
}
\DeclareAcronym{GW}{
  short = GW ,
  long = gravitational wave ,
  short-plural = s ,
}
\DeclareAcronym{GWB}{
  short = GWB ,
  long = gravitational wave background ,
  short-plural = s ,
}
\DeclareAcronym{SGWB}{
  short = SGWB ,
  long = stochastic gravitational-wave background ,
  short-plural = s ,
}
\DeclareAcronym{CBC}{
  short = CBC ,
  long = compact binary coalescence ,
  short-plural = s ,
}
\DeclareAcronym{BBH}{
  short = BBH ,
  long = binary black hole ,
  short-plural = s ,
}
\DeclareAcronym{PBH}{
  short = PBH ,
  long = primordial black hole ,
  short-plural = s ,
}
\DeclareAcronym{LIGO}{
  short =LIGO ,
  long = Laser Interferometer Gravitational-Wave Observatory ,
  short-plural = ,
}
\DeclareAcronym{LVK}{
  short = LVK ,
  long = {LIGO, Virgo and KAGRA} ,
  short-plural = ,
}
\DeclareAcronym{ET}{
	short = ET ,
	long  = Einstein Telescope,
  short-plural =  ,
}
\DeclareAcronym{CE}{
	short = CE ,
	long  = Cosmic Explorer,
  short-plural =  ,
}
\DeclareAcronym{LISA}{
	short = LISA ,
	long  = Laser Interferometer Space Antenna,
  short-plural =  ,
}
\DeclareAcronym{BBO}{
	short = BBO ,
	long  = big bang observer,
  short-plural =  ,
}
\DeclareAcronym{DECIGO}{
	short = DECIGO ,
	long  = Deci-hertz Interferometer Gravitational wave Observatory,
  short-plural =  ,
}
\DeclareAcronym{PTA}{
  short = PTA ,
  long = pulsar timing array ,
  short-plural = s ,
}
\DeclareAcronym{FRW}{
  short = FRW ,
  long = Friedman-Robertson-Walker ,
  short-plural =  ,
}
\DeclareAcronym{CMB}{
  short = CMB ,
  long = cosmic microwave background ,
  short-plural =  ,
}
\DeclareAcronym{SSR}{
  short = SSR ,
  long = sound speed resonance ,
  short-plural =  ,
}
\DeclareAcronym{SIGW}{
  short = SIGW ,
  long = scalar-induced gravitational wave ,
  short-plural = s ,
}
\DeclareAcronym{SKA}{
  short = SKA ,
  long =  Square Kilometer Array ,
  short-plural =  ,
}
\DeclareAcronym{NANOGrav}{
  short = NANOGrav ,
  long =  North American Nanohertz Observatory for Gravitational Waves ,
  short-plural =  ,
}
\DeclareAcronym{PNG}{
  short = PNG ,
  long =  primordial non-Gaussianity
 ,
  short-plural =  ,
}
\DeclareAcronym{MS}{
  short = MS ,
  long =  Mukhanov-Sasaki
 ,
}

\begin{document}

\title{Anisotropies in Scalar-Induced Gravitational-Wave Background from Inflaton-Curvaton Mixed Scenario with Sound Speed Resonance}

\author{Yan-Heng Yu}
\affiliation{Theoretical Physics Division, Institute of High Energy Physics, Chinese Academy of Sciences, 19B Yuquan Road, Shijingshan District, Beijing 100049, China}
\affiliation{School of Physics, University of Chinese Academy of Sciences, 19A Yuquan Road, Shijingshan District, Beijing 100049, China}

\author{Sai Wang}
\email{Correspondence author: wangsai@ihep.ac.cn}
\affiliation{Theoretical Physics Division, Institute of High Energy Physics, Chinese Academy of Sciences, 19B Yuquan Road, Shijingshan District, Beijing 100049, China}
\affiliation{School of Physics, University of Chinese Academy of Sciences, 19A Yuquan Road, Shijingshan District, Beijing 100049, China}

\begin{abstract}

We propose a new model to generate large anisotropies in the scalar-induced gravitational wave (SIGW) background via sound speed resonance in the inflaton-curvaton mixed scenario. Cosmological curvature perturbations are not only exponentially amplified at a resonant frequency, but also preserve significant non-Gaussianity of local type described by $f_{\mathrm{nl}}$. Besides a significant enhancement of energy-density fraction spectrum, large anisotropies in SIGWs can be generated, because of super-horizon modulations of the energy density due to existence of primordial non-Gaussianity. A reduced angular power spectrum $\tilde{C}_{\ell}$ could reach an amplitude of $[\ell(\ell+1)\tilde{C}_{\ell}]^{1/2} \sim 10^{-2}$, leading to potential measurements via planned gravitational-wave detectors such as DECIGO. The large anisotropies in SIGWs would serve as a powerful probe of the early universe, shedding new light on the inflationary dynamics, primordial non-Gaussianity, and primordial black hole dark matter.

\end{abstract}

\maketitle

\acresetall

{\textbf{Introduction.}}---Analogue to the study of \ac{CMB} \cite{Planck:2018vyg,Planck:2018jri}, the anisotropies in \ac{GWB} encode crucial information about the primordial universe, and further, enable us to directly detect the physics during the earlier stages beyond the observational capabilities of \ac{CMB}, because \acp{GW} can propagate almost freely since their generation \cite{Bartolo:2018igk,Flauger:2019cam}. In fact, the dynamics of the early universe has left imprints on cosmological \ac{GWB} and hence the anisotropies in \ac{GWB} serve as a powerful tool to explore this fundamental physics \cite{LISACosmologyWorkingGroup:2022kbp}. Though strong evidence for a \ac{GWB} have been reported \cite{NANOGrav:2023gor,EPTA:2023fyk,Reardon:2023gzh,Xu:2023wog}, distinguishing different origins of such a \ac{GWB} based solely on its energy-density fraction spectrum is challenging, since different physical scenarios might predict the same spectrum. However, the anisotropies in \ac{GWB} would be useful for differentiating among multiple astrophysical and cosmological sources via examining differences in their angular power spectra \cite{Bartolo:2019oiq,Bartolo:2019yeu,Bartolo:2019zvb,Li:2023qua,Li:2023xtl,Wang:2023ost,Dimastrogiovanni:2022eir,Contaldi:2016koz,Jenkins:2018kxc,Jenkins:2018uac,Jenkins:2019uzp,Jenkins:2019nks,Bertacca:2019fnt,Cusin:2017fwz,Cusin:2017mjm,Cusin:2018rsq,Cusin:2019jhg,Cusin:2019jpv,Wang:2021djr,Mukherjee:2019oma,Bavera:2021wmw,Bellomo:2021mer,Pitrou:2019rjz,Canas-Herrera:2019npr,Geller:2018mwu,Jenkins:2018nty,Jenkins:2018uac,Jenkins:2018nty,Kuroyanagi:2016ugi,Olmez:2011cg,Dimastrogiovanni:2021mfs,ValbusaDallArmi:2023nqn,Cui:2023dlo,Adshead:2020bji,Dimastrogiovanni:2019bfl,Jeong:2012df,Liu:2020mru,Li:2021iva,Domcke:2020xmn,Jinno:2021ury,Geller:2018mwu,Kumar:2021ffi,Racco:2022bwj,Bethke:2013aba,Bethke:2013vca,ValbusaDallArmi:2023ydl}. 

As inevitable byproducts of the evolution of linear cosmological curvature perturbations, \acp{SIGW} \cite{Ananda:2006af,Baumann:2007zm,Espinosa:2018eve,Kohri:2018awv,Mollerach:2003nq,Assadullahi:2009jc,Domenech:2021ztg} are considered as a natural cosmological source for the anisotropic \ac{GWB} \cite{Bartolo:2019zvb,ValbusaDallArmi:2020ifo,Dimastrogiovanni:2021mfs,LISACosmologyWorkingGroup:2022kbp,LISACosmologyWorkingGroup:2022jok,Unal:2020mts,Malhotra:2020ket,Carr:2020gox,Cui:2023dlo,ValbusaDallArmi:2023nqn,Li:2023qua,Li:2023xtl}, serving as a probe of the inflationary dynamics. \acp{SIGW} have attracted lots of interests for the relationship to many vital topics like \ac{PNG} \cite{Maldacena:2002vr,Bartolo:2004if,Allen:1987vq,Bartolo:2001cw,Acquaviva:2002ud,Bernardeau:2002jy,Chen:2006nt,Chen:2010xka,Choudhury:2023tcn}, which describes the deviation from the Gaussian statistics of primordial curvature perturbations and thus reflects the interactions of quantum fields during inflation. As shown in Refs.~\cite{Bartolo:2019zvb,Li:2023qua,Li:2023xtl,Wang:2023ost}, \ac{PNG} plays a key role in generating the anisotropies in \acp{SIGW} on large scales. It leads to couplings between the large- and small-scale curvature perturbations, resulting in the superhorizon modulation of the energy density of \acp{SIGW}, which is manifested as the initial inhomogeneities of \acp{SIGW}. Finally, these inhomogeneities lead to the anisotropies in \acp{SIGW}. Conversely, the latter can give insights on \ac{PNG}. 

Despite the importance of the anisotropies in \ac{GWB}, they have not been observed yet in present observations. Only upper limits have been placed on the angular power spectrum by the \ac{NANOGrav} \cite{NANOGrav:2023tcn}, and the \ac{LVK} \cite{LIGOScientific:2016nwa,LIGOScientific:2019gaw,KAGRA:2021mth}. Therefore, any cosmological scenario generating significantly anisotropic \ac{GWB} would attract great attention, since it would open a new window to study the early universe. 

In this work, we will propose a novel model to generate significantly anisotropic \acp{SIGW} by \ac{SSR} \cite{Cai:2018tuh,Cai:2019jah,Chen:2019zza,Chen:2020uhe} in the inflaton-curvaton mixed scenario \cite{Langlois:2004nn,Ferrer:2004nv,Langlois:2008vk,Fonseca:2012cj}.
As a nonlinear mechanism, \ac{SSR} is capable of enhancing primordial curvature perturbations on small scales, although some upper limits on the curvature power spectrum have been established \cite{Chluba:2019kpb,Jeong:2014gna,Nakama:2014vla,Inomata:2016uip,Allahverdi:2020bys,Gow:2020bzo,Franciolini:2022pav,Franciolini:2022tfm,Wang:2022nml}. 
Such an enhancement potentially leads to the production of a large abundance of \acp{PBH}, which are considered as a competitive candidate of dark matter \cite{Sasaki:2018dmp,Carr:2020xqk}. 
Meanwhile, the curvaton scenario \cite{Linde:1996gt,Lyth:2001nq,Enqvist:2001zp,Lyth:2002my,Moroi:2001ct,Bartolo:2003jx,Sasaki:2006kq,Huang:2008ze,Beltran:2008aa,Enqvist:2005pg,Malik:2006pm,Pi:2021dft} is a class of well-motivated multi-field inflationary models with potential to generate a large \ac{PNG}, which is also considered as a possible non-adiabatic source of anisotropic \acp{GW}  \cite{Malhotra:2022ply}. 
Comparing to the curvaton scenario, where the curvature perturbations from inflaton are negligible, the inflaton-curvaton mixed scenario incorporates the curvature perturbations from both the inflaton and the curvaton.
The inflaton-curvaton mixed scenario with \ac{SSR} was firstly studied in Ref.~\cite{Chen:2019zza}, in which the curvaton acts as a spectator scalar field and is assumed to undergo \ac{SSR} to create enhanced entropy perturbations during inflation. 
When the curvaton decays into the radiation after inflation, these perturbations convert into curvature perturbations with a peaked feature in primordial curvature power spectrum and a large \ac{PNG}, which are both crucial for generating significantly anisotropies in \acp{SIGW}. 
The proposed model could be realized by combining curvaton scenario with D-brane dynamics in string theory \cite{Silverstein:2003hf,Alishahiha:2004eh}, k-inflation \cite{Armendariz-Picon:1999hyi,Garriga:1999vw}, another coupled field with heavy modes integrated out in the effective field theory \cite{Achucarro:2010da,Achucarro:2013cva,Pi:2017gih}, and so on. 

We will also show that the anisotropies in \acp{SIGW}, with a reduced angular power spectrum $\ell(\ell+1)\,\Tilde{C}_\ell \sim 10^{-4}$, could be larger by five orders of magnitude than those in \ac{CMB}. This spectrum is potentially measured in future by multi-band \ac{GW} detectors, such as \ac{LVK} \cite{Harry_2010,VIRGO:2014yos,Somiya:2011np}, \ac{ET} \cite{Hild:2010id}, \ac{LISA} \cite{2019BAAS...51g..77T,Smith:2019wny}, {Taiji} \cite{Hu:2017mde}, \ac{SKA} \cite{Carilli:2004nx} and \ac{DECIGO} \cite{Seto:2001qf,Kawamura:2020pcg}. Once being observed, it would serve as a powerful probe of both \ac{PNG} and \acp{PBH}, and give implications for the inflationary dynamics.

{\textbf{Inflaton-curvaton mixed scenario with SSR.}}---We briefly review the inflaton-curvaton mixed scenario, following the conventions of Refs.~\cite{Langlois:2004nn,Ferrer:2004nv,Langlois:2008vk,Fonseca:2012cj}. Besides the inflaton $\phi$, there is a spectator field, i.e., the curvaton $\chi$, with its mass satisfying $m_\chi\ll H$, where $H$ is the Hubble parameter during inflation. The fields $\phi$ and $\chi$ are weakly coupled, and can be decomposed into their background components ($\overbar{\phi}$ and  $\overbar{\chi}$) and fluctuations ($\delta\phi$ and $\delta\chi$), with the assumption that these fluctuations are Gaussian. While $\delta\phi$ generates the curvature perturbations $\zeta_\phi$, $\delta\chi$ conveys the entropy perturbations $\cS_\chi$, which subsequently transform into the curvature perturbations $\zeta_\chi$ when $\chi$ decays into radiation after the end of inflation but before the start of primordial nucleosynthesis, as constrained by \ac{CMB} observations \cite{Planck:2018jri}. Therefore, the total curvature perturbations $\zeta_\mathrm{tot}$ are comprised of contributions from both ${\zeta_\phi}$ and ${\zeta_\chi}$. Though $\chi$ subdominates the energy density of the inflationary universe, ${\zeta_\chi}$ can be comparable to ${\zeta_\phi}$ or even dominant. 

In this model, $\delta\chi$ undergoes \ac{SSR} during inflation \cite{Chen:2019zza}. On spatially flat slices, the evolution of $\delta\chi_k$ obeys the \ac{MS} equation, i.e., \cite{Mukhanov:1988jd,Sasaki:1986hm}
\begin{equation}
    v''_k(\eta)+\left(c_s^2 k^2-\frac{z''}{z}\right)v_k(\eta)=0\ ,
\end{equation}
where we define $v_k= a\delta\chi_k/c_s$ and $z= a\overbar{\chi}'/(c_s \cH)$, a prime denotes a derivative with respect to the conformal time $\eta$, $c_s$ is the sound speed, $a$ is a scale factor of the universe, and $\cH$ is a conformal Hubble parameter. We suppose that $c_s$ oscillates with $\eta$, namely, \cite{Cai:2018tuh}
\begin{equation}\label{eq:cs2}
    c_s^2(\eta)=1-2\xi\left[1-\cos{(2k_0 \eta)}\right]\ ,
\end{equation}
where $\xi$ and $k_0$ stand for the oscillating amplitude and frequency, respectively. 
The oscillation begins at $\eta_i$, which is deep in the horizon, i.e., $-k_{0}\eta_i\gg 1$, and ends at the horizon exit $-k_{0}\eta_e\simeq1$. 
In the limit of slow-roll approximation and due to the temporal oscillation of $c_s$, the \ac{MS} equation can be recast into the Mattieu equation, which is a characteristic of \ac{SSR}, i.e., \cite{Chen:2019zza}
\begin{equation}\label{eq:Mattieu}
  \frac{\ud^2 v_k}{\ud x^2}
  +\left(A_k-2B_k\cos{2x}\right)v_k
  =0\ .
\end{equation}
Here, we introduce $x=-k_0\eta$, $A_k=(k^2/k_0^2)(1-2\xi)$, and $B_k=(2-k^2/k_0^2)\xi$, with an approximation of $\xi\ll1$. One of important features of the Mattieu equation is that $v_k$ (and equivalently $\delta\chi_k$) is exponentially amplified at a characteristic scale $k_\mathrm{S}\sim(1+\xi)k_0$ with resonant width $\sim \xi k_0$, indicating an amplification factor $|v_{k_\mathrm{S}}/v_{k\neq k_\mathrm{S}}|\sim e^{-\xi k_\mathrm{S} \eta/2}$ \cite{Cai:2018tuh,Chen:2019zza}. The amplified $\delta\chi$ finally leads to an enhancement of $\zeta_\mathrm{tot}$ at around $k_\mathrm{S}$, as showed in the following. 

{\textbf{Primordial curvature perturbations.}}---We investigate $\zeta_j$ ($j=\phi,\chi,\mathrm{tot}$) using $\delta N$ formalism \cite{Sasaki:1995aw,Starobinsky:1985ibc,Wands:2000dp,Sasaki:2006kq}. In this formalism, the superhorizon $\zeta_j$ on uniform-$\rho_\mathrm{tot}$ slices can be expressed as \cite{Lyth:2004gb,Sasaki:2006kq}, 
\begin{equation}\label{eq:zetaj}
  \zeta_j(t,\bx)
  =\delta N(t,\bx)+\frac{1}{3}\int_{\overbar{\rho}_j(t)}^{\rho_j(t,\bx)} \frac{\ud \rho'_j}{(1+w_j)\,\rho'_j}\ .
\end{equation}
Here, $\rho_j$ (and $\overbar{\rho}_j$) is the local (and background) energy density, $w_j$ is the equation-of-state parameter, and $\delta N$ is the local perturbation expansion from the initial spatially flat slice to the final uniform-$\rho_\mathrm{tot}$ slice. Specially, Eq.~(\ref{eq:zetaj}) naturally gives $\zeta_\mathrm{tot}=\delta N$ on a uniform-$\rho_\mathrm{tot}$ slice.

Firstly, we consider the generation of entropy perturbations $\cS_\chi=3(\zeta_\chi-\zeta_\phi)$ arising from $\delta\chi$ \cite{Lyth:2002my}. 
At the end of inflation, $\phi$ decays into radiation, which inherits $\rho_\phi$ and $\zeta_\phi$ from $\phi$. 
Since that time, both the radiation and $\chi$ are relativistic until $H$ decreases to $m_\chi$. 
Once $H\sim m_\chi$, $\chi$ begins to oscillate around the bottom of its potential and behaves as pressureless matter with $\rho_\chi=m_\chi^2 \chi^2_\mathrm{osc}$ \cite{Langlois:2008vk}, where $\chi_\mathrm{osc}$ represents the root-mean-square value of the oscillating curvaton. 
The uniform-$\rho_\mathrm{tot}$ slice at the beginning of the oscillation is characterized as $ \zeta_\phi=\delta N$, since $\rho_\chi$ is still subdominant \cite{Langlois:2008vk}. 
Eq.~(\ref{eq:zetaj}) gives $\rho_\chi=\overbar{\rho}_\chi\, e^{3(\zeta_\chi-\zeta_\phi)}=\overbar{\rho}_\chi\, e^{\cS_\chi}$ on this slice, leading to a relation between $\cS_\chi$ and $\delta\chi_{\mathrm{osc}}$ through $\rho_\chi$, i.e., 
\begin{equation}\label{eq:osc}
  m_\chi^2 \,(\overbar{\chi}_\mathrm{osc}+\delta\chi_\mathrm{osc})^2=m_\chi^2 \,\overbar{\chi}_\mathrm{osc}^2\, e^{\cS_\chi}\ .
\end{equation}
In this work, we assume that $\chi$ has a quadratic potential, which indicates that $\delta\chi_\mathrm{osc}/\,\overbar{\chi}_\mathrm{osc}=\delta\chi_{\ast}/\,\overbar{\chi}_{\ast}$ is a constant \cite{Lyth:2001nq,Lyth:2002my,Kohri:2012yw}, with ${}_\ast$ denoting the field value at horizon exit. 
With this assumption, we further connect $\cS_\chi$ to $\delta\chi_\ast$ by expanding Eq.~(\ref{eq:osc}) up to the second order, i.e.,
\begin{equation}\label{eq:S}
  \cS_\chi=2\,\frac{\delta \chi_\ast}{\overbar{\chi}_\ast}
  -\left(\frac{\delta \chi_\ast}{\overbar{\chi}_\ast}\right)^2
  =
  \cS_{\chi,g}-\frac{1}{4}\,\cS_{\chi,g}^2\ ,
\end{equation}
where we introduce $\cS_{\chi,g}= 2\,\delta \chi_\ast/\,\overbar{\chi}_\ast$ to represent a Gaussian component of $\cS_\chi$. 

Secondly, $\cS_\chi$ convert into adiabatic perturbations and contribute to $\zeta_\mathrm{tot}$. 
During the oscillation phase, $\rho_\chi/\rho_\mathrm{tot}$ continues to increase while $\cS_\chi$ remain constant. 
When $H$ drops to the decay rate of $\chi$, we assume that $\chi$ suddenly decays into radiation on a uniform-$\rho_\mathrm{tot}$ slice. 
This slice is characterized by $\delta N=\zeta_\mathrm{tot}$, where $\zeta_\mathrm{tot}$ is valued at the time right after the decay. 
Therefore, Eq.~(\ref{eq:zetaj}) implies the following relation \cite{Sasaki:2006kq}
\begin{equation}\label{eq:dec}
  \Omega_{\chi,\mathrm{d}}\,e^{3(\zeta_\chi-\zeta_\mathrm{tot})}+(1-\Omega_{\chi,\mathrm{d}})\,e^{4(\zeta_\phi-\zeta_\mathrm{tot})}=1\ ,
\end{equation}
with $\Omega_{\chi,\mathrm{d}}$ being the energy-density fraction of $\chi$ at the time right before the decay. 
Eq.~(\ref{eq:dec}) completely determines the relation between $\zeta_\mathrm{tot}$ and $\zeta_\chi$. 
We further expand Eq.~(\ref{eq:dec}) to the second order and express $\zeta_\mathrm{tot}$ in terms of $\cS_\chi$, namely,
\begin{eqnarray}\label{eq:zeta}
  \zeta_\mathrm{tot}
  &=&\zeta_\phi+\frac{r}{3}\,\cS_\chi+\frac{r(1-r)(3+r)}{18}\,\cS_\chi^2
  \nonumber\\
  &=&\zeta_\phi+\frac{r}{3}\,\cS_{\chi,g}+\frac{r}{18}\left(\frac{3}{2}-2r-r^2\right)\,\cS_{\chi,g}^2\ ,
\end{eqnarray}
where we define $r= 3\,\Omega_{\chi,\mathrm{d}}/(4-\Omega_{\chi,\mathrm{d}})$ with $0<r<1$. 
Through Eqs.~(\ref{eq:S}) and (\ref{eq:zeta}), the nonlinear evolution of $\delta \chi$ during inflation can be reflected on $\zeta_\mathrm{tot}$. 

Here, we investigate a dimensionless primordial power spectrum $\Delta^2_{\zeta_g}$ for the Gaussian curvature perturbations $\zeta_g$. 
In Eq.~(\ref{eq:zeta}), we assume that $\zeta_\phi$ is Gaussian, so the Gaussian component of $\zeta_\mathrm{tot}$ is $\zeta_g=\zeta_\phi+(r/3)\,\cS_{\chi,g}$. 
For convenience, we introduce $\lambda(k)$ to parameterize $\Delta^2_{\zeta_g}$ as  $\Delta^2_{\zeta_g}=\Delta^2_{\zeta_\phi}+\Delta^2_{\zeta_\chi}=[1+\lambda(k)]\,\Delta^2_{\zeta_\phi}$, where $\Delta^2_{\zeta_\phi}$ is a scale-invariant spectrum and $\lambda(k)\propto k\,|(v_k)_\ast|^2$ \cite{Chen:2019zza} is scale-dependent. 
On large scales related to $k_\mathrm{L}$, $\lambda$ is taken as a scale-invariant value, as denoted by $\lambda_\mathrm{L}$. 
On small scales at around $k_\mathrm{S}$, $\lambda$ is exponentially amplified and given by $\lambda_\mathrm{S} \sim \lambda_\mathrm{L}\,e^{-\xi k_\mathrm{S} \eta_i}\sim  \lambda_\mathrm{L}\,e^{\xi \, e^{\Delta N}}$ \cite{Chen:2019zza}, where $\Delta N$ is the e-folding number during \ac{SSR}. 
Therefore, we separate $\zeta_g$ as $\zeta_g=\zeta_{g\mathrm{L}}+\zeta_{g\mathrm{S}}$ \cite{Tada:2015noa}, where $\zeta_{g\mathrm{L}}$ is the large-scale component at $k_\mathrm{L}$ contributed from both $\zeta_\phi$ and $\cS_{\chi,g}$ of non-resonant modes, while $\zeta_{g\mathrm{S}}$ is the small-scale component at $k_\mathrm{S}$ arising from $\cS_{\chi,g}$ of resonant modes. 
Phenomenologically, we model $\Delta^2_{\zeta_g}$ as a scale-invariant background $\Delta^2_{\zeta_{g\mathrm{L}}}$ plus a resonant peak $\Delta^2_{\zeta_{g\mathrm{S}}}$ centered at $k_\mathrm{S}$, i.e., 
\begin{equation}\label{eq:Pls}
  \Delta^2_{\zeta_g}(k)
  =\Delta^2_{\zeta_{g\mathrm{L}}}+\Delta^2_{\zeta_{g\mathrm{S}}}
  =\cA_\mathrm{L}+
  \frac{\cA_{\mathrm{S}}}{\sqrt{2\pi}\sigma}\,e^{-\frac{1}{2\sigma^2}\ln^2{(k/k_\mathrm{S})}}\ ,
\end{equation}
where the large-scale spectral amplitude $\cA_\mathrm{L}$ is extrapolated from observations of \ac{CMB}, i.e., $\cA_\mathrm{L}\simeq2.1\times10^{-9}$ at a pivot scale $k_{\mathrm{p}}=0.05\,\mathrm{Mpc}^{-1}$ \cite{Planck:2018vyg}. 
On small scales, the spectral width $\sigma$ depends on the resonant width $\sim \xi k_0$. 
For example, we get $\sigma\sim10^{-2}$ when considering $\xi\sim 0.1$. 
The small-scale spectral amplitude $\cA_\mathrm{S}$ is determined by $\cA_\mathrm{S}/\cA_\mathrm{L}\sim(\xi/2)\, e^{\xi \, e^{\Delta N}}$. 
It describes the enhancement of $\zeta_g$ and could be closely related to the formation of \acp{PBH} \cite{Cai:2018tuh,Cai:2019jah,Chen:2019zza,Chen:2020uhe}. 
For example, $\Delta N \sim 5.4$ (or $10.2$) is enough for $\xi\sim 10^{-1}$ (or $10^{-3}$) to induce $\cA_\mathrm{S}\sim \cO(1)$, implying an efficient mechanism to produce abundant \acp{PBH}.

{\textbf{Primordial non-Gaussianity.}}---In the proposed model, the local-type \ac{PNG}, as denoted by $\fnl$, is not only large, but also scale-dependent. The non-Gaussian curvature perturbations $\zeta_{\mathrm{tot}}$ can be parameterized by the Gaussian one $\zeta_{g}$, i.e. \cite{Komatsu:2001rj}, 
\begin{equation}\label{eq:fnl def}
  \zeta_{\mathrm{tot},\bk}=\zeta_{g,\bk}+\frac{3}{5} \int \frac{\ud^3 \bp}{(2\pi)^{3/2}}\ f_\mathrm{nl}(k,p,|\bk-\bp|)\, \zeta_{g,\bk-\bp}\zeta_{g,\bp}\ .
\end{equation}  
Based on Eqs.~(\ref{eq:zeta}) and (\ref{eq:fnl def}), $f_\mathrm{nl}$ can be read as \cite{Langlois:2008vk,Fonseca:2012cj}
\begin{equation}\label{eq:fnl}
f_\mathrm{nl}(k_1,k_2,k_3)= \left(\frac{5}{4r}-\frac{5}{3}-\frac{5r}{6}\right)\alpha(k_1,k_2,k_3)\ ,
\end{equation}
where $\alpha= \Sigma'\,(r/3)^4 P_{\cS_{\chi,g}}(k_1)P_{\cS_{\chi,g}}(k_2)/\,\Sigma'\, P_{\zeta_g}(k_1)P_{\zeta_g}(k_2)$ with $P(k)=(2\pi^2/k^3)\Delta^2(k)$ being a dimensional power spectrum.
Here, we introduce a symbol $\Sigma'$ to represent $\Sigma'P(k_1)P(k_2)= P(k_1)P(k_2)+P(k_2)P(k_3)+P(k_3)P(k_1)$.

The \ac{PNG} $f_\mathrm{nl}$ in Eq.~(\ref{eq:fnl}) has two obvious features. 
One concerns that a large $f_\mathrm{nl}$ can be generated if $r$ is small, which implies that the nonlinear fluctuation should be more violent for a more subdominant $\chi$ to generate a certain amount of curvature perturbations. 
The other one is that $f_\mathrm{nl}$ is scale-dependent. 
Considering the scale dependence of $\lambda(k)$ and $P(k_\mathrm{L})\gg P(k_\mathrm{S})$, we can approximate the scale-dependent $\alpha(k_1,k_2,k_3)$ in Eq.~(\ref{eq:fnl}) in the squeezed limit $k_1\approx k_2\gg k_3$ as follows 
\begin{align}\label{eq:alpha}
\alpha \simeq
\left\{
\begin{aligned}
& \left(1+\lambda_\mathrm{L}^{-1}\right)^{-2},\ k_1,k_2,k_3\ \mathrm{at\ large\ scales}\ , \\
& \left(1+\lambda_\mathrm{S}^{-1}\right)^{-2},\ k_1,k_2,k_3\ \mathrm{at\ small\ scales}
\ , \\
& \left[\left(1+\lambda_\mathrm{S}^{-1}\right)\left(1+\lambda_\mathrm{L}^{-1}\right)\right]^{-1}, 
\end{aligned}
\right.
\\
\quad\quad k_1,k_2\ \mathrm{at\ small\ scales},\ k_3\ \mathrm{at\ large\ scales}\nonumber\ .
\end{align}
For simplicity, we denote $f_\mathrm{nl}$ in the three cases in Eq.~(\ref{eq:alpha}) as $f_\mathrm{nl,L}$, $f_\mathrm{nl,S}$ and $f_\mathrm{nl,LS}$, respectively. 
$f_\mathrm{nl,L}$ is constrained by the observations of \ac{CMB} as $f_\mathrm{nl,L}=-0.9\pm5.1$ at $68\%$ confidence level \cite{Planck:2019kim}, $f_\mathrm{nl,S}$ with $\lambda_\mathrm{S}\gg 1$ reduces to the standard result of the curvaton scenario \cite{Bartolo:2003jx,Sasaki:2006kq,Huang:2008ze,Beltran:2008aa}, while $f_\mathrm{nl,LS}$ describes the strength of nonlinear couplings between the large- and small-scale perturbations. 
As will be demonstrated, $f_\mathrm{nl,S}$ and $f_\mathrm{nl,LS}$ play key roles in the energy-density fraction spectrum and the angular power spectrum of \acp{SIGW}, respectively. 
They are constrained by measurements of $f_\mathrm{nl,L}$ via $f_\mathrm{nl,LS}^2\simeq f_\mathrm{nl,L}f_\mathrm{nl,S}$.

{\textbf{Energy-density fraction spectrum of SIGWs.}}---The energy density of \acp{SIGW} could be decomposed as $\rho_\mathrm{gw}(\eta,\bx)=\overbar{\rho}_\mathrm{gw}(\eta)+\delta\rho_\mathrm{gw}(\eta,\bx)$, where $\bx$ is the coarse-grained location depending on the finite angular resolution of \ac{GW} detectors, $\overbar{\rho}_\mathrm{gw}$ is the homogeneous and isotropic background, and $\delta\rho_\mathrm{gw}$ stands for the inhomogeneities on this background.
Corresponding to $\overbar{\rho}_\mathrm{gw}$ and $\delta\rho_\mathrm{gw}$, we define the energy-density fraction spectrum $\overbar{\Omega}_\uGW $ and the density contrast $\delta_\uGW$ of \acp{SIGW}, which are given by \cite{Maggiore:1999vm}
\begin{subequations}
\begin{align}
    \overbar{\rho}_\uGW (\eta) &= \rho_c(\eta) \int \ud \ln q\   \overbar{\Omega}_\uGW (\eta,q)\ ,\label{eq:Omega}\\
    \delta\rho_\uGW (\eta,\bx) &=\rho_c(\eta) \int \ud\ln{q}\  \ud^2 \hat{\bq}\,\, \frac{1}{4\pi}\,\overbar{\Omega}_\mathrm{gw}(\eta,q)\, \delta_\uGW (\eta,\bx,\bq)\ .\label{eq:delta}
\end{align}    
\end{subequations}
Here, we introduce $q=|\bq|$ and $\hat{\bq}=\bq/q$ with $\bq$ being the wavevector of \acp{SIGW}, and $\rho_c$ is the critical energy density of the universe at $\eta$. 
We focus solely on Eq.~(\ref{eq:Omega}) in this section while leave Eq.~(\ref{eq:delta}) to a study of the anisotropies in \acp{SIGW} in the next section.

At the production time $\eta_\mathrm{in}$ of \acp{SIGW}, the energy-density fraction spectrum is given by  $\overbar{\Omega}_\mathrm{gw}(\eta_\mathrm{in},q)\sim\langle\zeta_\mathrm{tot}^4\rangle$ \cite{Ananda:2006af,Baumann:2007zm}, where $\langle...\rangle$ stands for the spatial average. 
When evaluating $\overbar{\Omega}_\mathrm{gw}$, we can simply treat $\zeta_\mathrm{tot}$ as the small-scale curvature perturbations $\zeta_\mathrm{S}=\zeta_{g\mathrm{S}}+(3/5)\,f_\mathrm{nl,S}\, \zeta_{g\mathrm{S}}^2$, since contributions from the large-scale ones is negligible due to $\cA_\mathrm{L}\ll\cA_\mathrm{S}$. 
Therefore, for small scales of observational interest, we obtain $\overbar{\Omega}_\mathrm{gw}\sim\langle\zeta_\mathrm{S}^4\rangle $. Its complete analysis has been presented in Refs.~\cite{Adshead:2021hnm,Ragavendra:2021qdu,Abe:2022xur,Li:2023qua,Li:2023xtl}. 
Other related works can be found in Refs.~\cite{Cai:2018dig,Unal:2018yaa,Atal:2021jyo,Ragavendra:2020sop,Yuan:2020iwf,Yuan:2023ofl,Garcia-Saenz:2022tzu,Zhang:2021rqs,Domenech:2017ems,Garcia-Bellido:2017aan,Nakama:2016gzw}.

\begin{figure}[h]
\includegraphics[width=\linewidth]{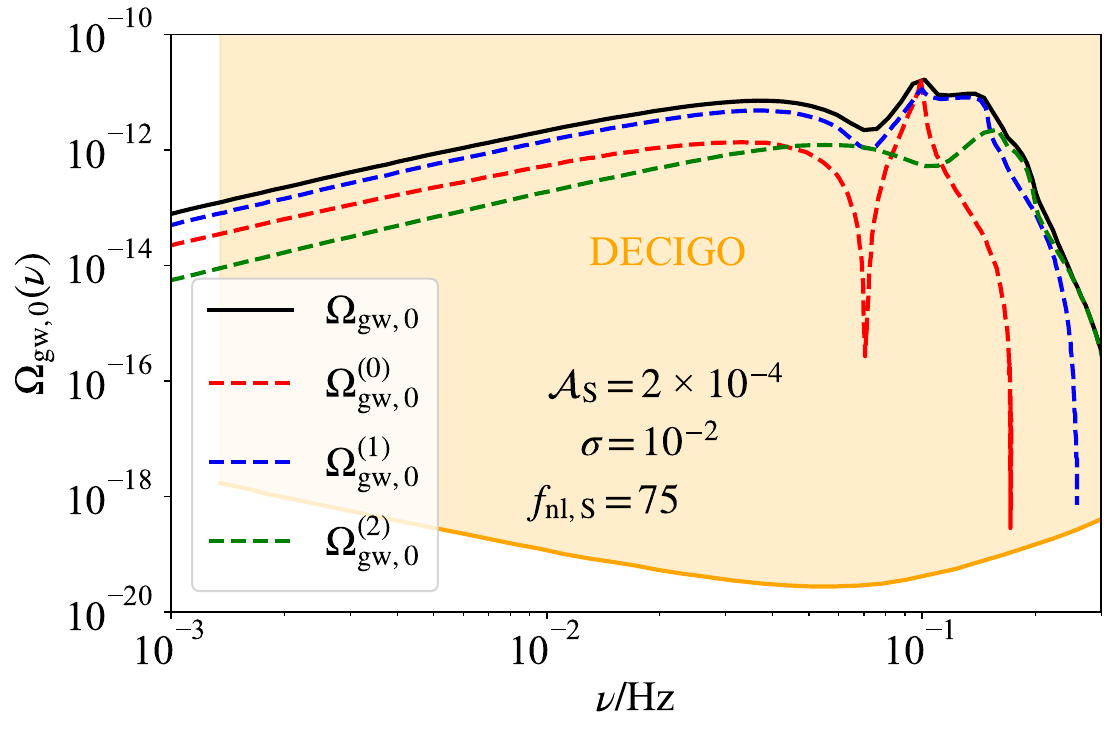}
\caption{Energy-density fraction spectrum $\overbar{\Omega}_{\uGW,0} (\nu)$ (black solid curve) and its components $\overbar{\Omega}_\mathrm{gw,0}^{(0)}$ (red dashed curve), $\overbar{\Omega}_\mathrm{gw,0}^{(1)}$ (blue dashed curve), and $\overbar{\Omega}_\mathrm{gw,0}^{(2)}$
(green dashed curve) at the frequency $\nu_\mathrm{peak}=0.1$Hz. For comparison, the sensitivity of \ac{DECIGO} is displayed as the orange shaded region.}\label{fig:Omega0}
\end{figure}

The present-day energy-density fraction spectrum of \acp{SIGW} is given by \cite{Wang:2019kaf} 
\begin{equation}
   \overbar{\Omega}_{\uGW,0} (\nu) \simeq \Omega_{\mathrm{rad}, 0}\, \overbar{\Omega}_\uGW (\eta_\mathrm{in},q)\big|_{q=2\pi \nu} \ ,
\end{equation}
where the present-day physical energy-density fraction of radiation is $h^2\Omega_{\mathrm{rad},0}=4.2\times 10^{-5}$ with the dimensionless Hubble constant $h=0.674$  \cite{Planck:2018vyg}. 
For simplicity, we denote the $\cO(f^{2m}_\mathrm{nl,S})$ component of $\overbar{\Omega}_\mathrm{gw,0}$ as $\overbar{\Omega}_\mathrm{gw,0}^{(m)}$ $(m=0,1,2)$.
In Fig.~\ref{fig:Omega0}, we depict $\overbar{\Omega}_{\uGW,0}$ and $\overbar{\Omega}_\mathrm{gw,0}^{(m)}$ with variable $\nu=q/(2\pi)$, and compare them with the sensitivity curve of \ac{DECIGO} \cite{Braglia:2021fxn}. 
It is shown that $\overbar{\Omega}_{\uGW,0}$ is enhanced at around $\nu_{\mathrm{peak}}= (2/\sqrt{3})(k_\mathrm{S}/2\pi)$, and a large $f_\mathrm{nl,S}$ enlarges the spectral amplitude and obviously changes the spectral index.

{\textbf{Angular power spectrum of SIGWs.}}---Following Refs.~\cite{Bartolo:2019zvb,Li:2023qua,Li:2023xtl}, we show that in Eq.~(\ref{eq:delta}), the large-scale $\delta\rho_\mathrm{gw}$ leads to the anisotropies in \acp{SIGW}.    
Here, we disregard the small-scale inhomogeneities, because the horizon at $(\eta_\mathrm{in},\bx_{\mathrm{in}})$ is extremely small for the present-day observer at $(\eta_0,\bx_0)$, and the observed energy density along a line-of-sight is an average over a quantity of such horizons \cite{Bartolo:2019zvb,Li:2023qua,Li:2023xtl,Wang:2023ost}.
The observed large-scale $\delta_\uGW(\eta_0,\bx_0,\bq)$ in Eq.~(\ref{eq:delta}) is contributed by both the initial density contrast and the Sachs-Wolfe effect \cite{Sachs:1967er}, i.e., 
\begin{equation}\label{eq:delta0}
    \delta_\uGW(\eta_0,\bx_0,\bq)
    =\delta_\uGW(\eta_\mathrm{in},\bx_\mathrm{in},\bq)
    +\left[4-n_\mathrm{gw,0}(\nu)\right]
    \Phi(\eta_\mathrm{in},\bx_\mathrm{in})
    \ ,
\end{equation}
where we define $n_\mathrm{gw,0}(\nu)=\partial\ln{\overbar{\Omega}_\mathrm{gw,0}(\nu)}/\,{\partial\ln{\nu}}$, and the Bardeen potential follows $\Phi(\eta_\mathrm{in},\bx_\mathrm{in})\propto\zeta_{g\mathrm{L}}$. 
Compared with the Sachs-Wolfe effect, the integrated Sachs-Wolfe effect is negligible \cite{Bartolo:2019zvb}. 

In Eq.~(\ref{eq:delta0}), we focus on the term $\delta_\uGW(\eta_\mathrm{in},\bx_\mathrm{in},\bq)$, which arises from $f_\mathrm{nl,LS}$, indicating the couplings between $\zeta_{g\mathrm{L}}$ and $\zeta_{g\mathrm{S}}$. 
Such couplings can spatially modulate the energy density of \acp{SIGW}, which were produced by $\zeta_\mathrm{S}$, on superhorizon scales. 
To demonstrate this picture, we expand $\overbar{\Omega}_\mathrm{gw}\sim\langle\zeta_\mathrm{tot}^4\rangle$ according to the order of $\zeta_{g\mathrm{L}}$, i.e.,
\begin{equation}
    \langle\zeta_\mathrm{tot}^4\rangle
    \sim
    \langle\zeta_\mathrm{S}^4\rangle
    +\cO(\zeta_{g\mathrm{L}})\,f_\mathrm{nl,LS}\,\langle \zeta_{g\mathrm{S}}\,\zeta_\mathrm{S}^3 \rangle 
    +\cO(\zeta^2_{g\mathrm{L}})\ .
\end{equation}
The large-scale $\delta_\uGW(\eta_\mathrm{in},\bx_\mathrm{in},\bq)$ can be obtained from the $\cO(\zeta_{g\mathrm{L}})$ component of $\langle\zeta_\mathrm{tot}^4\rangle$, leading to $\delta_\uGW(\eta_\mathrm{in},\bx_\mathrm{in},\bq)\sim \cO(\zeta_{g\mathrm{L}})\,f_\mathrm{nl,LS}\,\langle \zeta_{g\mathrm{S}}\,\zeta_\mathrm{S}^3 \rangle$ at the first order \cite{Bartolo:2019zvb,Li:2023qua,Li:2023xtl,Wang:2023ost}. 
This formula implies that a large amplitude of $f_\mathrm{nl,LS}$ leads to significant initial inhomogeneities in \acp{SIGW}. 
Moreover, higher-order components of $\langle\zeta_\mathrm{tot}^4\rangle$ are negligible due to $\cA_{\mathrm{L}}\ll  \cA_{\mathrm{S}}$.

The observed anisotropies in \acp{SIGW}, as characterized by a reduced angular power spectrum $\Tilde{C}_\ell$, arise from the large-scale $\delta_\uGW(\eta_0,\bx_0,\bq)$.
With the assumption of the cosmological principle, we define $\Tilde{C}_\ell$ as follows 
\begin{equation}
   \left\langle\delta_{\uGW,0,\ell m} (q)\, \delta_{\uGW,0,\ell' m'}^\ast (q)\right\rangle \Big|_{q=2\pi\nu}=\delta_{\ell \ell'}\delta_{mm'} \Tilde{C}_\ell (\nu)\ , 
\end{equation}
where $\delta_{\uGW,0,\ell m}(q)$ represents the coefficients of the spherical harmonic expansion of $\delta_\uGW(\eta_0,\bx_0,\bq)$, i.e.,
\begin{equation}
  \delta_\uGW(\eta_0,\bx_0,\bq)=\sum_{\ell m}\ \delta_{\uGW,0,\ell m}(q) Y_{\ell m}(\hat{\bq})\ .
\end{equation}
Though two $\zeta_{g\mathrm{S}}$ with a large angular separation are uncorrelated, $\zeta_{g\mathrm{L}}$ serves as a ``bridge'' to correlate them. 
Hence, we get $\Tilde{C}_\ell\sim\langle\delta^2_\uGW(\eta_0,\bx_0,\bq)\rangle\propto\langle\zeta_{g\mathrm{L}}^2\rangle\sim \Delta^2_{\zeta_{g\mathrm{L}}}$.
Following a diagrammatic approach, Ref.~\cite{Li:2023qua} presented comprehensive calculations of $\Tilde{C}_{\ell}$.
The result is given as
\begin{equation}\label{eq:Cl}
  \Tilde{C}_{\ell}(\nu)  =\frac{18\pi\,\cA_{{\mathrm{L}}}}{25\,\ell(\ell+1)}\left[f_\mathrm{nl,LS}\,\frac{\Omega_\mathrm{ng,0}}{\overbar{\Omega}_\mathrm{gw,0}}+\left(4-n_\mathrm{gw,0}\right)\right]^2\ ,
\end{equation}
where we introduce $\Omega_\mathrm{ng,0}=2^3\, \overbar{\Omega}^{(0)}_\mathrm{gw,0}+2^2\, \overbar{\Omega}^{(1)}_\mathrm{gw,0}$ as a shorthand notation. 
The first and second terms in the square brackets inherit from Eq.~(\ref{eq:delta0}) the initial inhomogeneities and the Sachs-Wolfe effect, respectively.

\begin{figure}[h]
\includegraphics[width=\linewidth]{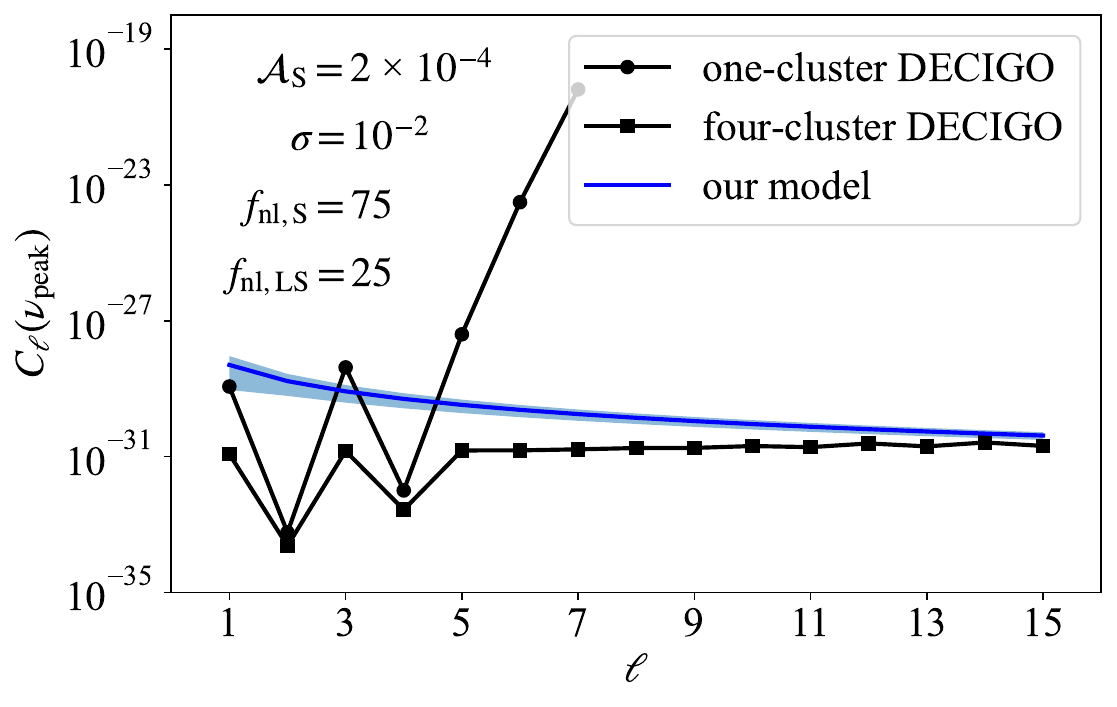}
\caption{Angular power spectrum anticipated by our model (blue curve) versus the sensitivity curves of \ac{DECIGO} (black curves) at the frequency $\nu_\mathrm{peak}=0.1$Hz. 
The black circles and squares represent the noise angular power spectra at each multipole $\ell$ for one-cluster and four-cluster DECIGO, respectively \cite{Capurri:2022lze}.
Shaded blue region signifies an inevitable uncertainty at 68\% confidence level due to the cosmic variance, i.e., $\Delta C_\ell/C_\ell=\sqrt{2/(2\ell+1)}$. }\label{fig:DECIGO}
\end{figure}

We demonstrate that large anisotropies in \acp{SIGW} can be generated in our model. 
We consider that the energy density fraction of the curvaton is small at the time of its decay and the curvature perturbations from curvaton are comparable to those of inflation at $\sim k_\mathrm{L}$ while dominant at $\sim k_\mathrm{S}$. 
Specifically, we quantify these as $\{r=0.017,\, \lambda_\mathrm{L}=0.5,\, \lambda_\mathrm{S}\gg 1\}$, resulting in \ac{PNG} $\{f_\mathrm{nl,S}=75,\, f_\mathrm{nl,LS}=25,\, f_\mathrm{nl,L}=8.3\}$ through Eqs.~(\ref{eq:fnl}--\ref{eq:alpha}). 
We further quantify the \ac{SSR} parameters as $\{\xi\simeq 0.12,\, \Delta N \simeq 4.8\}$, leading to $\{\cA_\mathrm{S}= 2\times10^{-4},\,\sigma=10^{-2}\}$ in Eq.~(\ref{eq:Pls}), which are within the constraints from various observations \cite{Cai:2018tuh}. 
Therefore, we anticipate the reduced angular power spectrum at around $\nu_\mathrm{peak}$ to be 
\begin{equation}\label{eq:Cl10-2}    
\sqrt{\ell(\ell+1)\Tilde{C}_\ell(\nu_\mathrm{peak})}\sim 10^{-2}\ .
\end{equation}
Regarding observations, we define an angular power spectrum as $C_\ell(\nu)=\left[\,\overbar{\Omega}_\mathrm{gw,0}(\nu)/4\pi\,\right]^2 \Tilde{C}_{\ell}(\nu)$. 
It could reach $\ell(\ell+1)C_\ell\sim 10^{-28}$, which is potentially observable for future \ac{GW} detectors like \ac{DECIGO} \cite{Seto:2001qf,Kawamura:2020pcg}. 
In Fig.~\ref{fig:DECIGO}, we depict the anticipated $C_\ell$ at $\nu_\mathrm{peak}=$ 0.1Hz and compare it with the sensitivity curve of \ac{DECIGO} in the same frequency band.
Moreover, we show an inevitable uncertainty (68\% confidence level) due to the cosmic variance in a blue shaded region.

While the $C_\ell$ above may not be sufficiently large for detection by \ac{LISA}, adjusting the parameters in our model could potentially yield an observable angular power spectrum, say, $C_\ell\sim 10^{-23}$ at around $\nu_\mathrm{peak}= 1\mathrm{mHz}$, for two LISA-like clusters \cite{Capurri:2022lze}. 
However, the corresponding $\Tilde{C}_\ell$ may not be so large. 
The maximum achievable anisotropies would only be around $[\ell(\ell+1)\Tilde{C}_\ell]^{1/2}\sim 10^{-3}$, where the parameters could be set as $\{\cA_\mathrm{S}= 8\times10^{-3},\,\sigma=10^{-2},\,f_\mathrm{nl,S}=15,\, f_\mathrm{nl,LS}=12,\, f_\mathrm{nl,L}=10\}$.

{\textbf{Conclusions and discussion.}}---In this work, we showed that the large anisotropies in \acp{SIGW} are generated when the curvaton undergoes \ac{SSR} in the inflaton-curvaton mixed scenario. 
Introduced by \ac{SSR}, the exponentially amplified curvature perturbations could induce the enhanced energy-density fraction spectrum of \acp{SIGW} around the resonant frequency. 
The large \ac{PNG} in this model leads to a large-scale modulation of the energy density of \acp{SIGW}, resulting in the large anisotropies, i.e., $[\ell(\ell+1)\Tilde{C}_\ell]^{1/2}\sim 10^{-2}$. 
These anisotropies have potential to be detected by future \ac{GW} detectors like DECIGO.

We may use both the energy-density fraction spectrum $\overbar{\Omega}_{\uGW,0}$ and the angular power spectrum $C_\ell$ of \acp{SIGW} as complementary probes to study the dynamics of the early universe.
The peaked feature in both spectra might serve as a characteristic \ac{GW} signature of this model. The peak width and frequency encode the amplitude $\xi$ and mode $k_0$ of the oscillation of $c_s^2$, respectively, while the peak amplitude says the duration of \ac{SSR}, characterized by $\Delta N$. Additionally, the imprints of $f_\mathrm{nl,S}$ and $f_\mathrm{nl,LS}$ in $\overbar{\Omega}_{\uGW,0}$ and $C_\ell$ enable us to explore the parameter space of $\{r,\,\lambda_\mathrm{L},\,\lambda_\mathrm{S}\}$, which carries vital information about the contributions from curvaton to energy density and curvature perturbations. The measurement of all these parameters is helpful in testing various inflationary models. 

Different from the existing studies in Refs.~\cite{Li:2023qua}, which assumed the scale independence of $f_\mathrm{nl}$, our present work considered the scale dependence of $f_\mathrm{nl}$. 
The anisotropies in \acp{SIGW} offer a remarkable method to measure $f_\mathrm{nl,LS}$, which reveals underlying physics of the nonlinear couplings between long- and short-wavelength modes. 
These anisotropies may initiate a lot of interests in the model-building studies related with the scale-dependent \ac{PNG}. 
When higher-order \ac{PNG} such as $g_\mathrm{nl}$ is incorporated \cite{Li:2023xtl}, our result in Eq.~(\ref{eq:Cl10-2}) would not change too much due to $|g_\mathrm{nl}|\ll f_\mathrm{nl}^2$ when $r\ll1$ in this model \cite{Sasaki:2006kq,Fonseca:2012cj}.

The proposed model can lead to efficient production of \acp{PBH}, shedding new insights on the nature of cold dark matter. 
Considering the large positive $f_\mathrm{nl,S}$, we roughly estimate the abundance of \acp{PBH} as \cite{Byrnes:2012yx,Meng:2022ixx}
\begin{eqnarray}\label{eq:pbh}
  f_\mathrm{PBH}
  &\simeq &2.5\times 10^{8}\  \left(\frac{g_{\ast,\mathrm{f}}}{10.75}\right)^{-\frac{1}{4}} \left(\frac{\nu}{5.0\,\mathrm{nHz}}\right) 
  \nonumber\\
  && \times\frac{1}{2} \  \mathrm{erfc}\left(\frac{\sqrt{1+4\,\zeta_c\,(3f_\mathrm{nl,S}/5)}-1}{2\,(3f_\mathrm{nl,S}/5)\times\sqrt{2\cA_\mathrm{S}}}\right)\ ,
\end{eqnarray}
where $g_{\ast,\mathrm{f}}$ stands for the number of relativistic degrees of freedom at the formation time of \acp{PBH}, $\zeta_c$ is the critical curvature perturbation for gravitational collapse, and $\mathrm{erfc}(x)$ denotes the complementary error function with variable $x$. 
If considering $g_{\ast,\mathrm{f}}\sim100$ \cite{Saikawa:2018rcs} and $\zeta_c=0.7$ \cite{Green:2004wb}, we can obtain $f_\mathrm{PBH}\sim 1$ at $\nu\sim0.1$Hz with the parameters adopted in this work.
All of cold dark matter could be accounted for by these abundant \acp{PBH} in the asteroid mass range, which has not been explored by the present observations \cite{Carr:2020gox}.
However, we should note that the above estimation is sensitively model-dependent. 
For instance, altering the value of $\zeta_c$, considering the $f_\mathrm{nl}$-dependence of $\zeta_c$, or employing other methods based on the critical density contrast $\delta_c$ could potentially impact the resulting $f_\mathrm{PBH}$.
 
Our model could also be extended into more general cases.
For example, a different expression for $c_s^2$ oscillation from Eq.~(\ref{eq:cs2}) will not change our framework to obtain an enhanced $\cA_\mathrm{S}$. Additionally, alternative resonant mechanisms, such as an oscillating inflaton potential ~\cite{Cai:2019bmk,Zhou:2020kkf,Cai:2021yvq,Peng:2021zon,Cai:2021wzd,Inomata:2022yte,Meng:2022low}, could also be considered. 
Furthermore, if considering a non-quadratic potential for $\chi$, the evolution of $\chi$ between the horizon exit and oscillating phase would necessitate numerical investigation. 
While this could lead to some modifications in $f_\mathrm{nl}$, the picture of generating large $f_\mathrm{nl}$ with a small $r$ will remain unchanged \cite{Sasaki:2006kq,Fonseca:2012cj}.

\begin{acknowledgments}
We appreciate Mr. Jun-Peng Li and Dr. Shi Pi for helpful discussion. 
This work is partially supported by the National Natural Science Foundation of China (Grant No. 12175243), the National Key R\&D Program of China No. 2023YFC2206403, the Science Research Grants from the China Manned Space Project with No. CMS-CSST-2021-B01, and the Key Research Program of the Chinese Academy of Sciences (Grant No. XDPB15).
\end{acknowledgments}




\bibliography{SSR}

\end{document}